\documentclass[12pt]{article}

\usepackage{amsmath}
\usepackage{amssymb}
\usepackage{amsfonts}
\usepackage{eucal}
\usepackage{epsfig}

\newcommand{\beal}{\begin{subequations} \begin{align}}
\newcommand{\eeal}{\end{align} \end{subequations}}
\newcommand{\be}{\begin{equation}}
\newcommand{\bea}{\begin{eqnarray}}
\newcommand{\eea}{\end{eqnarray}}
\newcommand{\ba}{\begin{array}}
\newcommand{\ea}{\end{array}}
\newcommand{\ee}{\end{equation}}

\makeatletter
\@addtoreset{equation}{section}
\renewcommand{\theequation}{\thesection.\arabic{equation}}
\makeatother
\expandafter\ifx\csname mathbbm\endcsname\relax

\else

\fi
\def\hmu{\hat{\mu}}
\def\hnu{\hat{\nu}}

\def\super{$PSU(2|2)\times PSU(2|2)\times U(1)$ }
\begin{document}
\baselineskip 18pt

\begin{titlepage} 
\hfill
\vbox{
    \halign{#\hfil         \cr
           SU-ITP-03/31\cr
           SLAC-PUB-10255\cr
           hep-th/0312155 \cr
           } 
      }  
\vspace*{6mm}
\begin{center}  
{\Large {\bf
Giant Hedge-Hogs: \\
Spikes on Giant Gravitons}}
\vspace*{5mm}
\vspace*{1mm}

{\bf Darius Sadri${}^{1,2}$ and  M. M. Sheikh-Jabbari$^{1}$}

\vspace*{0.4cm}
{\it ${}^1${Department of Physics, Stanford University\\
382 via Pueblo Mall, Stanford CA 94305-4060, USA}}

\vspace*{0.2cm}

{\it ${}^2${Stanford Linear Accelerator Center,
Stanford  CA 94309, USA }}\\

\vspace*{0.4cm}

{\tt darius@itp.stanford.edu,$\,$jabbari@itp.stanford.edu}

\vspace*{1cm}
\end{center}

\begin{center}
{\bf\large Abstract}
\end{center}

We consider giant gravitons on the maximally supersymmetric plane-wave background of type IIB
string theory. Fixing the light-cone gauge, we deduce the low energy effective light-cone 
Hamiltonian of the three-sphere giant graviton.
At first order, this is a $U(1)$ gauge theory on $\mathbb{R} \times S^3$. 
We place sources in this  effective gauge theory.
Although non-vanishing net electric charge configurations are disallowed by Gauss' law, electric 
dipoles can be formed. {}From the string theory point of view these dipoles can be understood as 
open strings piercing the three-sphere, generalizing the usual BIons to the giant gravitons (BIGGons). 
Our results can be used to give a two dimensional (worldsheet) description of
giant gravitons, similar to Polchinski's description for the usual D-branes, in agreement with the 
discussions of hep-th/0204196.


\end{titlepage}


\section{Introduction}

Giant gravitons \cite{McGreevy:2000cw} were first discussed in the context of
$m-2$-branes moving on the sphere in an $AdS_n \times S^m$ background, where it was
observed that such particles blow up inside $S^m$, losing their point-like structure, and where
their size was related to their angular momentum.
In fact they are branes which  couple to  background form fields as dipoles, in contrast to
the manner in which flat branes couple to form fields; they carry zero net form field charge, but 
a non-vanishing dipole moment. It is this dipole coupling that is responsible for their blowing up.
In the $AdS_5 \times S^5$ background, they are three dimensional branes.
Initial interest in these objects arose from their connection to non-commutative physics,
and the scaling of their size with angular momentum was recognized as a hallmark
of non-commutativity. The giant gravitons preserve the same supercharges as the
graviton multiplet \cite{Grisaru:2000zn}, and are $1/2$ BPS, forming short representations
of the superalgebra.

Giant gravitons which expand into the $AdS$ part of the space-time have also
been constructed, and they carry the same quantum numbers as sphere giant gravitons
\cite{Grisaru:2000zn,Hashimoto:2000zp}.
The vibration spectrum of small fluctuations for giant gravitons which have expanded in
either the $AdS$ or the sphere directions have been studied \cite{Das:2000st}, where
it was found that the masses were independent of the radius and angular momentum
of the giant graviton, depending only on the curvature scale of the background.

BPS solutions of type IIB supergravity describing giant gravitons carrying angular
momentum along the sphere are available \cite{Myers:2001aq}, and collections of
giant gravitons act as external sources which give rise to extremal limits of charged black holes (superstars), where the horizon coincides with the singularity (which is hence naked) in the $AdS$ component of the space-time.
Solutions of eleven dimensional supergravity (again BPS) characterizing giant gravitons on
$AdS_7 \times S^4$ and $AdS_4 \times S^7$ appeared in \cite{Leblond:2001gn},
where they also found to contain naked singularities, sourced by giant gravitons 
interpreted as spherical $M2$ and $M5$ branes.

The $AdS/CFT$ duality suggests that giant gravitons should correspond to some
operators in a dual conformal field theory, where these operators are chiral
primary. The dual operators (for both sphere and AdS giants) have been constructed
\cite{Balasubramanian:2001nh, Corley:2001zk, Balasubramanian:2002sa, 
Ouyang:2002vg, Berenstein:2003ah},
and some correlation functions have also been computed.
The sphere giant gravitons correspond to operators constructed from determinants and
sub-determinants of the scalar fields in the $\mathcal{N}=4$ super Yang-Mills theory (e.g. see 
footnote \ref{subdeterminant}). 
The determinants are associated with maximum size giant gravitons, carrying the
maximum angular momentum on the $S^5$. Similarly there have been proposals for the operators dual 
to giant gravitons inside $AdS$ \cite{Corley:2001zk}.


A new maximally supersymmetric type IIB supergravity solution (``the'' plane-wave), arising as the 
Penrose limit of $AdS_5 \times S^5$ \cite{Blau:2001ne}, has attracted much interest 
in the literature, largely because of its connection to the $AdS/CFT$ duality
\cite{BMN}, and the fact that
the Green-Schwarz superstring action, in light-cone gauge, is exactly solvable
\cite{Metsaev:2001bj,Metsaev:2002re}.
This Plane-wave/super-Yang-Mills duality is a specification of the usual $AdS/CFT$
correspondence in the Penrose limit;
it states that strings on a plane-wave background are dual to a particular large R-charge 
sector of $\mathcal{N}=4, \ D=4$ superconformal $U(N)$ gauge theory.\footnote{
A review can be found in \cite{Sadri:2003pr}.}
The study of giant gravitons was then extended from $AdS_5 \times S^5$ to the plane-wave,
and interesting issues stemming from the nature of the dual operators were addressed
\cite{Balasubramanian:2002sa,Takayanagi:2002nv}, in particular the question of
open strings in the dual gauge theory.

In a different line of pursuit, Callan and Maldacena \& Gibbons \cite{Callan:1997kz, 
Gibbons:1997xz} considered the low energy effective theory for a single brane, which gives rise 
via the Dirac-Born-Infeld action, to a $U(1)$ gauge theory, and showed that electric point charges 
could be interpreted as end-points of fundamental strings ending on the brane, and the dual 
magnetic charges as D-strings similarly ending on the brane. They demonstrated the profile these 
strings took and showed that they are in fact BPS solutions.
These BPS solutions of the linearized (Maxwell) equations
match the solutions of the full non-linear Born-Infeld theory equations, hence called BIons.
A similar setup was argued to hold for other BPS brane junctions, giving
a characterization in terms of local configurations of fields in the effective
description of the brane in terms of a gauge theory.

We study giant gravitons on the plane-wave appearing as the Penrose limit of $AdS_5 \times S^5$,
with a particular focus on the behaviour of charges in the worldvolume gauge theory
and their interpretation in terms of open strings. We find solutions which generalize the
usual BIons to giant gravitons, allowing an open string world-sheet description of such
giant gravitons. The outline of our paper follows:
In section \ref{GG-in-plane-wave} we present the low energy effective theory
describing giant gravitons on the plane-wave, working in light-cone gauge.
We find two zero-energy configurations (vacua), corresponding to a zero-size
giant graviton and one of finite size, with radius given in terms of the string
coupling $g_s$, the light-cone momentum $p^+$, and a scale $\mu$ for measuring
energies. We then analyze the spectrum of fluctuations, writing
their eigenfrequencies and eigenmodes. We find agreement between the physical modes
and those of $\mathcal{N}=4$ super-Yang-Mills on $\mathbb{R} \times S^3$.
Higher order corrections are studied, extracting the effective coupling of the
theory, and the relation of this coupling to the dual BMN gauge theory parameters is
discussed. In section \ref{gauge-theory-on-GGs}, we analyze the behaviour of the
worldvolume theory when gauge fields are turned on, presenting the spectrum of the
gauge field.
In section \ref{biggons}, we turn our attention to the BIon solutions on giant
gravitons (BIGGons), explicitly solving for the scalar and gauge field configurations, and
interpret them via energy considerations as fundamental strings piercing the
giant gravitons. The supersymmetry and stability of the configurations is also
addressed.
In a final section, we summarize our conclusions and outline possible
future directions for pursuit.
An appendix is included, summarizing the harmonics which appear in the main
body of the paper.

\section{Giant Gravitons in the Plane-Wave Background}\label{Lightcone-Hamiltonian}
\label{GG-in-plane-wave}

In this section we focus on the $3+1$ dimensional Dirac-Born-Infeld action in the plane-wave
background. First we note that due to symmetries of the plane-wave background, in particular
translational symmetry along the light-like directions $x^+$ and $x^-$ \cite{Sadri:2003pr} (similar
to the case of strings on the same background \cite{Metsaev:2001bj}), fixing the light-cone gauge
will simplify considerably the action. The zero energy solutions to the light-cone Hamiltonian in
the sector with light-cone momentum $\mu p^+$ is a sphere of radius $R^2=\mu p^+ g_s$. This sphere
is a giant graviton \cite{McGreevy:2000cw}. It is worth noting that fixing the light-cone gauge, in
the language of rotating (orbiting) branes of \cite{McGreevy:2000cw, Hashimoto:2000zp}, corresponds
to going to the rest frame of the giant graviton.

We also study fluctuation modes of the giant gravitons by expanding the light-cone Hamiltonian 
about
the zero energy solutions. The frequencies of these modes, as in the case of giant gravitons on the
$AdS_5\times S^5$ background \cite{Das:2000st}, are independent of their radius. We next turn on
the fermions and work out the full fermionic terms of the light-cone Hamiltonian and the
frequencies of their small fluctuations. We also briefly discuss higher order interaction terms in
the Hamiltonian and the fact that they may be analyzed in a systematic perturbation expansion with
the effective coupling $g_{eff}$ \eqref{coupling}.

\subsection{Low energy effective dynamics in light-cone gauge}\label{LCgauge-fixing}

The low energy effective action for a D-brane is
\be
  S = S_{DBI} + S_{CS} \, ,
\ee
with the Dirac-Born-Infeld action
\be \label{DBI-action}
  S_{DBI} = - T_p \int d^{p+1} \zeta \:   e^{-\phi} \:
  \sqrt{-det \left( G_{\hmu\hnu} +B_{\hmu\hnu} + F_{\hmu\hnu} \right)} \, ,
\ee
where hatted Greek indices are used for the worldvolume 
coordinates ranging from zero to $p$. We have set $2 \pi\alpha^\prime=1$; factors of $\alpha'$ can be 
reintroduced on dimensional grounds when necessary. 
We will consider D3-branes, for which $p=3$ and the dilaton background
is constant, in which case $g_s=e^{\phi}$. We first consider the case
where, in addition to the constant dilaton, only the metric is turned on,
and drop (consistently) the other forms. The gauge field $F_{\hmu\hnu}$, however, would be 
considered in section \ref{gauge-theory-on-GGs}.
Our metric conventions are those of Polchinski \cite{Book};
we work with a mostly plus metric for the worldvolume and target space.
Note that the physical tension for this D-brane is $T_p/g_s$.
The Chern-Simons term describing the coupling to the background RR four form is
\be
  S_{CS} = q \int C_4 \, ,
\ee
with $q$ the charge of the brane. For BPS configurations
the charge and tension are equal.
$G_{\hmu\hnu}$ is the pullback of the space-time metric onto the worldvolume of the
brane, and $C_4$ is the pullback of the RR four-form. They are given by
\be \label{pullback-metric}
  G_{\hmu\hnu} = \partial_{\hmu} X^\mu \partial_{\hnu} X^\nu g_{\mu \nu}
\ee
and
\be
  C_4 = \frac{1}{4!} \:
  \left( \partial_{\hmu_0} X^{\mu_0} \cdots  \partial_{\hmu_3} X^{\mu_3} \:
  C_{\mu_0 ... \mu_3} \right) \:
  d\zeta^{\hmu_0} \wedge \cdots \wedge d\zeta^{\hmu_3} \, .
\ee
The $X^\mu$ give the embedding coordinates of the brane in the target
space-time, i.e. $\mu=0,\cdots, 9$,
and $\zeta^{\hmu}$ are local coordinates on the brane worldvolume.
$C_{\mu_0 ... \mu_3}$ is the space-time RR four-form coupling to the
worldvolume.

We are working in a background specified by the maximally supersymmetric type IIB plane-wave  (here we will follow the notation and conventions of \cite{Sadri:2003pr})
\begin{subequations}\label{background}
\begin{align}
  ds^2  =  -2 dx^+ dx^- & -\mu^2(x^i x^i + x^a x^a) {(dx^+)}^2 + dx^i dx^i+ dx^a dx^a \, ,\\
  F_{+ijkl} &= \frac{4}{g_s} \mu\ \epsilon_{ijkl} \, ,\ \ \ \ \ \ 
  F_{+abcd}=\frac{4}{g_s} \mu \ \epsilon_{abcd} \, .
\end{align}
\end{subequations}
{}From (\ref{background}b) it is easy to read off the RR four-form potential $C$, as $F=dC$, and 
we have
\be\label{fourform}
  C_{+ijk} = -\frac{\mu}{ g_s} \epsilon_{ijkl} x^l \, , \ \ \ \ 
  C_{+a b c} = -\frac{\mu}{ g_s} \epsilon_{a b c d} x^{d} \, ,
\ee
which has the virtue of maintaining the translational symmetry along $x^+$.
We have chosen our coordinates to make manifest the $SO(4) \times SO(4)$ symmetry
of the transverse directions, labeling the two $SO(4)$'s with
$i,j=1,2,3,4\, ;\ \ a,b,c,d=5,6,7,8$. For a more detailed discussion on the isometries of the background we refer the reader to \cite{Sadri:2003pr}.

We separate the space and time indices on the brane worldvolume as
$\zeta=(\tau=\sigma^0,\sigma^r)$, with $p,q,r=1,2,3,$ the space indices.
We will fix the light-cone gauge, setting
\be
  X^+ = \tau \, .
\ee
In order to ensure that the above solution for  $X^+$ is maintained by the dynamics, we should use a part of the gauge symmetries of the DBI action, which are the area preserving diffeomorphisms on the brane worldvolume, to set
\be\label{G0r}
  G_{0 r} = - \partial_r X^- + \partial_\tau X^I \partial_r X^I= 0 \, .
\ee
We have used upper-case indices to denote all eight transverse coordinates, where 
$I=(i,a)=1,2,\cdots,8$.

Next we note that the background \eqref{background} is $X^-$ independent, (it is a cyclic
coordinate), and hence the momentum 
conjugate to $X^-$, the light-cone momentum $p^+$, is a constant of motion:
\bea\label{pplus}
p^+=-\frac{\partial {\cal L}}{\partial{\partial_\tau X^-}}&=&
- \frac{1}{g_s}G^{00}\sqrt{-det\ G}\cr
&=& - \frac{1}{g_s}\sqrt{\frac{-det\ G_{rs}}{G_{00}}} \, .
\eea
To obtain \eqref{pplus} we have used the fact that $G_{0 r}=0$ implies $G^{00}=1/G_{00}$.
The light-cone Hamiltonian $P^-$ (i.e. momentum conjugate to $X^+$)  is then found to be
\bea\label{Hlc}
P^-\equiv-\frac{\partial {\cal L}}{\partial{\partial_\tau X^+}}&=&
- \frac{1}{g_s}G^{00}\sqrt{-det\ G}(\partial_\tau X^- + \mu^2 X^I X^I)-
\frac{1}{6}\epsilon^{rps}
C_{+IJK}\partial_r X^I\partial_p X^J\partial_s X^K \cr
&=& p^+(\partial_\tau X^- + \mu^2 X^I X^I)-\frac{1}{6}\epsilon^{rps}C_{+IJK}\partial_r 
X^I\partial_p 
X^J\partial_s X^K.
\eea
Using \eqref{pplus} we can solve $G_{00}$ and hence $\partial_\tau X^-$ for $p^+$ and $det 
\: G_{rs}$:
\be\label{dtauX-}
  G_{00} = -2 \partial_\tau X^- -\mu^2 X^I X^I +
  \partial_\tau X^I \partial_\tau X^I =
  - \frac{det G_{rs}}{(p^+g_s)^2} \, .
\ee
Inserting $\partial_\tau X^-$ from \eqref{dtauX-} into the light-cone Hamiltonian and noting
that the momenta conjugate to $X^I$ are
\be\label{PI}
P_I=\frac{\partial {\cal L}}{\partial{\partial_\tau X^I}}=p^+\partial_\tau X^I \, ,
\ee
we obtain the light-cone Hamiltonian density
\be\label{Hlightcone}
  {H}_{l.c.} = \frac{1}{2 p^+}P^I P^I+ V(X^i,X^a) \, ,
\ee
where
\bea\label{potential}
  V(X^i,X^a)&=&\frac{\mu^2 p^+}{2}(X_i^2+X_a^2)+\frac{1}{2 p^+g_s^2} det \: G_{rs}\cr
  &-&\frac{\mu}{6g_s} \Bigl(
  \epsilon^{i j k l} X^i\{ X^j, X^k, X^l \}+ \epsilon^{a b c d} X^a\{ X^b, X^c, X^d \} \Big) \, .
\eea
In the above, 
\[
G_{rs}=\partial_r X^i\partial_s X^i+\partial_r X^a\partial_s X^a \, ,
\]
and the brackets are ``Nambu brackets'' defined as
\be\label{Nambubracket}
  \{ F , G , K \} = \epsilon^{p q r} \partial_p F \partial_q G \partial_r K \, ,
\ee
where the antisymmetrization is with respect to worldvolume coordinates.
It is worth noting that as a result of light-cone gauge fixing the square-root in the DBI action 
has disappeared (see \eqref{potential}). This will help us perform a more detailed analysis of the 
light-cone Hamiltonian.
We should also keep in mind that in the light-cone gauge,
$\partial_r X^-$ are totally determined in terms of $X^I$ through \eqref{G0r}, i.e.
\be\label{constraint}
  -P_I \partial_r X^I + p^+ \partial_r X^- \approx 0\ ,
\ee
where $\approx$ is the ``weak'' equality, meaning that \eqref{constraint} should hold on the 
solutions of the equations of motion of the light-cone Hamiltonian.

\subsection{Zero energy configurations}\label{vacuum-solutions}

We now search for classical minima of the light-cone Hamiltonian, and 
expand the potential $V(X^i,X^a)$ around these vacua to find the spectrum of small fluctuations 
about the 
vacua. First we note that if we set $X^a=0$, then
\be
det\ G_{rs}=det (\partial_r X^i\partial_s X^i)=\frac{1}{3!}\{ X^i, X^j, X^k \}\{ X^i, X^j, X^k \}
\, ,
\ee
and hence the potential becomes a perfect square
\be
V(X^i, X^a=0)=\frac{1}{2p^+}\left(\mu p^+ X^i-\frac{1}{6g_s}\epsilon_{ijkl}\{ X^j, X^k, X^l\}\right)^2.
\ee
The above potential has a minimum at   
\be\label{minima}
\mu p^+g_s \epsilon_{ijkl}X^l=\{ X^i, X^j, X^k\} \, .
\ee
Eq.\eqref{minima} has two solutions, one is the ``trivial'' vacuum, $X^i=0$, and the other one is a three-sphere of radius $R$, where\footnote{
Note that all lengths are measured in units of the string scale $\alpha'$. Recovering the $\alpha'$-factors, we have
$R^2/\alpha'=(\mu p^+\alpha') g_s$.} \cite{BMN} 
\be\label{radius}
R^2=\mu p^+ g_s
\ee
In other words if we set
\be\label{threesphere}
X^i=R x^i,\ \ \ \ \ \sum_{i=1}^4 x^2_i=1\ ,
\ee
it is easy to check that $\epsilon_{ijkl}x^i=\{x^j,x^k,x^l\}$. 
This three sphere is a giant graviton.

Both of these vacua are zero energy configurations. 
We could have easily found another minimum (zero energy configuration) corresponding to a three 
sphere grown in the $X^a$ directions sitting at $X^i=0$. Note also that both of these vacua are 
$1/2$ BPS; they annihilate all the dynamical supercharges of the background. In other words, all the fermionic generators of the $PSU(2|2)\times PSU(2|2)\times U(1)$ superalgebra
would kill these states.\footnote{Although both of these vacua are BPS, it has been argued that
the $X=0$ vacuum might be unstable under certain quantum corrections 
\cite{McGreevy:2000cw, Hashimoto:2000zp}. These arguments were finally confirmed for the case of 
spherical membranes and fivebranes in M-theory in a detailed analysis of the Matrix theory
describing M-theory on the maximally supersymmetric eleven dimensional plane-wave background, the BMN matrix theory. It has been shown that the $X=0$ vacuum for the membrane/fivebrane case is in fact a finite size fivebrane/membrane \cite{Maldacena:2002rb}. For the case of spherical three-branes, to the authors' knowledge the issue is not yet fully answered.}

\subsection{Spectrum about the vacua}\label{bosonic-spectrum}

We now study the spectrum of small fluctuations about these vacua. To do so, we expand the theory
about the vacua to second order in fluctuations.

\subsubsection{Spectrum about $X=0$ vacuum}

In this case the $det G_{rs}$ and the bracket terms would not  contribute to the quadratic 
Hamiltonian; they appear in the interactions, and the quadratic parts of the Hamiltonian are
\be\label{bosonic-X=0}
H_{X=0}^{(2)}= \frac{1}{2 p^+}P_i P_i+\frac{1}{2 p^+}P_a P_a+\frac{\mu^2 p^+}{2}X_i^2+\frac{\mu^2 
p^+}{2}X_a^2\ .
\ee
Therefore, there are eight modes, all with frequency $\mu$, that is the modes are particles of mass $\mu$. Of course for a generic low energy state one may excite many of these modes. 

\subsubsection{Spectrum about the three-sphere  vacuum}\label{X=R-spectrum}
 
If we parameterize the small fluctuations in the $X^i$ directions by $Y^i$, i.e. $X^i=R x^i + Y^i$, 
the quadratic Hamiltonian becomes
\bea \label{quad:Hamiltonian}
 H^{(2)}_{X=R} = \frac{1}{2 p^+}P_i P_i+\frac{1}{2 p^+}P_a P_a&+&
\frac{1}{2p^+} \left(\mu p^+ Y_i-\frac{R^2}{2g_s}\epsilon_{ijkl}\{x^j,x^k,Y^l\}\right)^2\cr
&+&\frac{1}{2p^+}\left( (\mu p^+)^2 X_a^2+\frac{R^4}{g_s^2} \partial_r X^a\partial_s X^a 
g_0^{rs}\right) \ ,
\eea
where $g_0^{rs}$ is the inverse of the metric on a unit three-sphere.
The bracket can be used to obtain generators of $SO(4)$ rotations along the 
three-sphere, explicitly: 
\be \label{rot-gens}
  \mathcal{L}_{ij} \Phi \equiv (x_j\partial_i-x_i\partial_j)\Phi= -\frac{1}{2} \epsilon_{ijkl} \{ 
x^k, x^l, \Phi \}\ .
\ee
In terms of ${\cal L}_{ij}$ the Hamiltonian takes a simple form   
\bea\label{H2X=R}
 H^{(2)}_{X=R} = \frac{1}{2 p^+}P_i P_i+\frac{1}{2 p^+}P_a P_a+
\frac{1}{2}\mu^2 p^+ (Y_i+{\cal L}_{ij}Y^j)^2+\frac{1}{2}\mu^2 p^+( 
X_a^2- \frac{1}{2} X^a {\cal L}_{ij}{\cal L}^{ij}X^a). 
\eea
The normal modes for the $Y^i$ directions about this vacuum satisfy the eigenvalue equation
\[
  \mathcal{L}_{ij} Y^j = \lambda Y_i \, ,
\]
with masses given by
\be
  M^2 = \mu^2 \left( 1 + \lambda \right)^2 \, .
\ee
The eigenvectors are vector spherical harmonics of the form
\begin{subequations}\label{eigenvectors-1}
\begin{align}
  Y^i_l &= S_{i i_1 \cdots i_l} \: x^{i_1} \cdots x^{i_l} \, , \ \ \ \ \ \ \ \ \ \ \ \ \qquad\qquad\qquad 
l\geq 0\\
  \tilde{Y}^i_l &= x^i \tilde{S}_{i_1 \cdots i_{l-1}} \: x^{i_1} \cdots x^{i_{l-1}} \:
  - \frac{l-1}{2l} \tilde{S}_{i i_1 \cdots i_{l-2}} x^{i_1} \cdots x^{i_{l-2}} \ , \quad l\geq 1,
\end{align}
\end{subequations}
where $S$ and $\tilde{S}$ are symmetric traceless $SO(4)$ tensors.
The eigenvectors (\ref{eigenvectors-1}a,b) correspond to the eigenvalues $\lambda = l,-(l+2)$, 
respectively. Both of these modes, although in different $SO(4)$ representations, would have the same mass:
\be
M_i=\mu (l+1) .
\ee
Physically these two modes correspond to geometric fluctuations of the brane in the radial
directions.

Of the modes describing the five directions $X^-,Y^i$, there remain three zero modes
\be \label{eigenvectors-2}
  \hat{Y}^i_l = A^i_{i_1 \cdots i_l} \: x^{i_1} \cdots x^{i_{l}} \, \quad l\geq 1,
\ee
where $A$ is symmetric in all lower indices and antisymmetric in the first upper and
first lower index (and hence  $x_i\hat{Y}^i_l=0$ for any $l$). These eigenvectors are associated with the eigenvalue $\lambda = -1$.
These zero modes are not  physical and correspond to gauge degrees of freedom associated with the shape preserving diffeomorphisms on the three-sphere.

The masses for $X^a$ fluctuations can be easily obtained, noting that the eigenvalues for the 
$SO(4)$ Casimir ${\cal L}^2$, are $-2l(l+2)$, with the corresponding $X^a$
\be\label{SO(4)_a}
X^a_l=S^a_{i_1\cdots i_l}x^{i_1}\cdots x^{i_l}\ , \quad \quad l\geq 0.
\ee
The masses are
\be
M^2_a=\mu^2[l(l+2)+1]=\mu^2(l+1)^2\ .
\ee

As we see, all the modes, $Y^i$'s and $X^a$'s, have the same mass. This is a direct result of the 
supersymmetry algebra of this background, which as discussed in \cite{Sadri:2003pr} is 
$PSU(2|2)\times 
PSU(2|2)\times U(1)$, and the fact that the light-cone Hamiltonian commutes with the supercharges; 
as a result all the states in the same supermultiplet should have the same mass.
This is in contrast with the eleven dimensional plane-wave superalgebra 
\cite{Dasgupta:2002ru}.\footnote{The $X^i=Rx^i$ vacuum, being a $1/2$ BPS state,
should, in the
dual ${\cal N}=4,\ D=4$ gauge theory, be represented by a chiral primary operator. And in 
our case, since we are working in the plane-wave background, it should be a BMN \cite{BMN} type 
operator. The corresponding operators have been introduced and studied in 
\cite{Balasubramanian:2001nh, Corley:2001zk, Balasubramanian:2002sa, Takayanagi:2002nv, 
Berenstein:2003ah}.
Let $Z^i_j,\ i,j=1,2,\cdots, N$, be one of the three complex scalar fields of an ${\cal N}=4,\ 
D=4,\ U(N)$ gauge theory (for more on conventions and notations see \cite{Sadri:2003pr}). Then 
\cite{Balasubramanian:2001nh}
\be\label{ggsph}
{\cal O}_J^{{S^5}}={\cal N}_J\frac{1}{J!(N-J)!} 
\epsilon_{{i_1}{i_2}\cdots{i_J}{k_{J+1}}\cdots {k_{N}}}\ 
\epsilon^{{j_1}{j_2}\cdots{j_J}{k_{J+1}}\cdots {k_{N}}}
Z^{i_1}_{j_1}Z^{i_2}_{j_2}\cdots Z^{i_J}_{j_J}\ ,
\ee
is the operator dual to a giant 
graviton grown in the $S^5$ direction
(the normalization factor ${\cal N}_J^2=\frac{(N-J)!}{N!}$ is chosen so that 
$\langle {\cal O}_J^{{S^5}} \bar{{\cal O}}_J^{{S^5}}\rangle=1$).
\[
{\cal O}_J^{{AdS}}=\frac{1}{J!}\sum_{\sigma\in {\cal S}_J}\ 
Z^{i_1}_{i_{\sigma(1)}}Z^{i_2}_{i_{\sigma(2)}}\cdots Z^{i_J}_{i_{\sigma(J)}}\ ,
\]
with ${\cal S}_J$ being the permutation group of length $J$, is proposed to 
describe giant gravitons grown in the $AdS$ directions \cite{Corley:2001zk}.
Note that in the plane-wave case (after the Penrose limit) the two giant gravitons grown in $S^5$
and in $AdS$ essentially become indistinguishable.

Using the above operators one may construct the dual gauge theory operators corresponding to the 
fluctuation modes of the giant graviton we have studied here. This can be done by insertion of 
``impurities'' in the sequence of $Z$'s, much like what has been done in \cite{BMN} for strings.
For example, if $\phi_a,\ a=1,2,3,4$ denote the other four scalars 
of the ${\cal N}=4,\ 
D=4$ gauge multiplet, then the dual gauge theory operators for $l=0, 1$ states of \eqref{SO(4)_a}
are \cite{Balasubramanian:2002sa, Berenstein:2003ah}
\[
{\cal O}_J^{{X^a_{l=0}}}={\cal N}_{J+1}\frac{1}{(J+1)!(N-J-1)!} 
\epsilon_{{i_1}{i_2}\cdots{i_{J+1}}{k_{J+2}}\cdots {k_{N}}}\
\epsilon^{{j_1}{j_2}\cdots{j_{J+1}}{k_{J+2}}\cdots {k_{N}}}
{(\phi^a)}^{i_1}_{j_1}Z^{i_2}_{j_2}Z^{i_3}_{j_3}\cdots Z^{i_{J+1}}_{j_{J+1}}\ ,
\]
\[
{\cal O}_J^{{X^a_{l=1}}}={\cal N}_{J+2}\frac{1}{(J+2)!(N-J-2)!} 
\epsilon_{{i_1}{i_2}\cdots{i_{J+2}}{k_{J+3}}\cdots {k_{N}}}
\epsilon^{{j_1}{j_2}\cdots{j_{J+2}}{k_{J+3}}\cdots {k_{N}}}
{(\phi^a)}^{i_1}_{j_1}Z^{i_2}_{j_2}Z^{i_3}_{j_3}\cdots {(\phi^b)}^{i_l}_{j_l}\cdots 
Z^{i_{J+2}}_{j_{J+2}}\ .
\]
Clearly these operators have $\Delta-J=1,2$, respectively. Similarly,
one may construct higher $l$ excitations by more insertions of $\phi$'s.
\label{subdeterminant}}

\subsubsection{Fermionic modes}\label{fermionic-modes}

For completeness we also work out the spectrum of fluctuations for the fermionic modes 
about both vacua.
The fermionic contributions to the DBI and CS parts of the action can be found using superspace 
coset 
techniques which make the superalgebra manifest (see for example \cite{Kallosh:1998zx}, and
\cite{Metsaev:2002re}, and also \cite{Maldacena:2002rb,Dasgupta:2002hx} for a similar treatment for
the case of the membrane and fivebrane on the eleven dimensional maximally supersymmetric
plane-wave).
After fixing $\kappa$ symmetry in light-cone gauge \cite{Metsaev:2002sg},
the new contributions to the
potential in the Hamiltonian are quadratic in the fermions, and are given by
\be \label{potential-fermion}
\begin{split}
  V^\psi =
   &\mu \psi^\dagger {}^{\alpha \beta} \psi_{\alpha \beta}+ 
\frac{2}{p^+ g_s} \biggl(
  \psi^\dagger {}^{\alpha \beta} (\sigma^{ij})_\alpha^{\: \: \: \delta} \:
  \{ X^i, X^j, \psi_{\delta \beta} \} +
  \psi^\dagger {}^{\alpha \beta} (\sigma^{ab})_\alpha^{\: \: \: \delta} \:
  \{ X^a, X^b, \psi_{\delta \beta} \}\biggr)+\\ 
&\mu\psi^\dagger {}^{\dot\alpha \dot\beta} \psi_{\dot\alpha \dot\beta} 
+\frac{2}{p^+ g_s} \biggl( 
\psi^\dagger {}^{\dot\alpha \dot\beta} 
(\sigma^{ij})_{\dot\alpha}^{\: \: \: \dot\delta} \:
  \{ X^i, X^j, \psi_{\dot\delta \dot\beta} \}+
\psi^\dagger {}^{\dot\alpha \dot\beta} 
(\sigma^{ab})_{\dot\alpha}^{\: \: \: \dot\delta} \:
  \{ X^a, X^b, \psi_{\dot\delta \dot\beta} \}
  \biggr) .
\end{split}
\ee
We have chosen to decompose the $SO(4) \times SO(4)$ fermions in terms of representations
of the two $SU(2)$'s appearing in each $SO(4)$, following the notation of \cite{Sadri:2003pr},
making manifest the fermion representations under the $PSU(2|2) \times PSU(2|2)$ part of the
superalgebra of the maximally supersymmetric plane-wave we are considering.
The explicit mass terms in $V^\psi$ for the fermions come from
a shift in $G_{\tau \tau}$ arising from fermionic contributions to the super-vielbein and the 
terms involving Nambu brackets from the Chern-Simons terms.
Following the notation of \cite{Sadri:2003pr}, the fermions $\psi$ are
spinors of two different $SU(2)'s$, one coming from the decomposition of each of the
two $SO(4)$'s into $SU(2) \times SU(2)$; in other words, $\psi$ above carries
two spinor indices (sitting in the same chirality representations). 
More details on our spinor conventions can be found in \cite{Sadri:2003pr}.
Note that in the potential \eqref{potential-fermion}, the two sets of fermions
with dotted and undotted indices do not couple to each other.

Expanding the potential around the three-sphere solution, setting
$X^i=R x^i$ and $X^a=0$, the quadratic part of the potential \eqref{potential-fermion} 
becomes, after using \eqref{radius} and \eqref{rot-gens}
\be \label{fermion-potential-quad}
\begin{split}
V^\psi_{(2)} = 
  \: & \mu \left(\psi^\dagger {}^{\alpha \beta}\psi_{\alpha \beta}-
  \psi^\dagger {}^{\alpha \beta} (\sigma^{ij})_\alpha^{\: \: \: \delta} \:
  \epsilon^{ijkl} \mathcal{L}_{kl} \psi_{\delta \beta} \right)+\\
 &\mu\left(\psi^\dagger {}^{\dot\alpha \dot\beta} \psi_{\dot\alpha \dot\beta}-
\psi^\dagger {}^{\dot\alpha \dot\beta} 
(\sigma^{ij})_{\dot\alpha}^{\: \: \: \dot\delta} \:
  \epsilon^{ijkl} \mathcal{L}_{kl} \psi_{\dot\delta \dot\beta}
  \right).
\end{split}
\ee
The spectrum of small fluctuations around this vacuum is given by solutions of the 
eigenvalue equation
\be \label{spinor-efs}
  \epsilon^{ijkl} (\sigma^{ij})_\alpha^{\: \: \: \beta} \:
  \mathcal{L}_{kl} \psi_\beta =
  \lambda \: \psi_\alpha \, ,
\ee
with similar equations for the other modes.
We have for clarity suppressed one of the indices since it is a bystander in the eigenvalue
equation.

The frequencies (masses) are then given by
\be
  \omega = \mu | 1 - \lambda | \, ,
\ee
where $\lambda$ is the eigenvalue corresponding to the excitation mode.
The eigenfunctions and corresponding eigenvalues (suppressing the inactive spinor index) are
\be
\begin{split}
  \psi^l_\alpha &= \left( \theta_{\alpha i_1 \ldots i_l} +
  \epsilon^{j i_1 k l}
  (\sigma^{kl})_\alpha^{\: \: \: \beta} \theta_{\beta j i_2 \ldots i_l} \right)
  x^{i_1} \cdots x^{i_l} \ \ \ \ \ \ \ \ \ \ \ \ \ \ \ \ \ \ \  \ \ \lambda = - l \, , \\
  \tilde{\psi}^l_\alpha &= \left(
  l \: \theta_{\alpha i_1 \ldots i_l} + (l+2)
  \epsilon^{i_1 j k l}
  (\sigma^{kl})_\alpha^{\: \: \: \beta}
  \theta_{\beta j i_2 \ldots i_l}
  \right)
  x^{i_1} \cdots x^{i_l} \ \ \ \ \ \ \ \ \ \ \ \lambda = l + 2 \, ,
\end{split}
\ee
where $l\geq 0$ and $\theta$, carrying the spinorial index, forms a totally symmetric traceless 
representation of $SO(4)$ in the
indices $j,i_1,\ldots,i_l$.\footnote{
It is straightforward but tedious to check that these are indeed eigenfunctions of
\eqref{spinor-efs}, making use of the identity
\[
\begin{split}
  (\sigma^{ij})_\alpha^{\: \: \: \beta} (\sigma^{kl})_\beta^{\: \: \: \rho} =
  &- \frac{1}{4} \Bigg[
  \delta_\alpha^\rho \left( \delta^{ik} \delta^{jl} - \delta^{il} \delta^{jk}
  + i \epsilon^{ijkl} \right) \\
  &+ 2 \left(
  \delta^{ik} (\sigma^{jl})_\alpha^{\: \: \:\rho} +
  \delta^{jl} (\sigma^{ik})_\alpha^{\: \: \:\rho} -
  \delta^{il} (\sigma^{jk})_\alpha^{\: \: \:\rho} -
  \delta^{jk} (\sigma^{il})_\alpha^{\: \: \:\rho}
  \right) \Bigg].
\end{split}
\]}
Therefore both $\tilde{\psi}^l_\alpha $ and ${\psi}^l_\alpha$ excitations have the same mass, 
$|\omega|$, equal to $\mu(l+1)$. As it is clear from \eqref{fermion-potential-quad},
fermions $\psi_{\dot\alpha\dot\beta}$ would also have the same mass. Hence all the 
bosonic and fermionic excitations about the $X^i=Rx^i$ vacuum $(Y^i_l,\tilde{Y}^i_l, X^a_l;
{\psi}^l_{\alpha\beta},\tilde{\psi}^l_{\alpha\beta},
{\psi}^l_{\dot\alpha\dot\beta},\tilde{\psi}^l_{\dot\alpha\dot\beta})$ have the same mass, as
expected from the $PSU(2|2)\times PSU(2|2)\times U(1)$ superalgebra, and fall into the same 
supermultiplet. However, these modes do not complete the multiplet (as there are 
two more fermions than bosons). These two extra bosonic modes correspond to a $U(1)$ gauge field 
living on the giant graviton; we will come back to this point in  section 
\ref{photon-spectrum-section}.

The frequencies of small perturbations around the zero size giant graviton are
simply given by $\mu$, arising from the explicit mass term in the potential.
The masses are then the same as the bosonic spectrum. In the $X=0$ vacuum, as opposed to the 
spherical vacuum, the number of scalar and fermionic excitations are both eight, i.e. 
there are no gauge field modes. 

Finally we would like to point out that, although we do not explicitly  show it here,
all the modes, about both vacua, fall into a BPS (short) multiplet of the $PSU(2|2)\times 
PSU(2|2)\times U(1)$ superalgebra. A study of representation theory of this superalgebra is an 
interesting open problem in need of a thorough analysis.

\subsection{Interaction terms}\label{effective-coupling}

So far we have only considered the quadratic terms around each of the two vacua. One may study the
theory perturbatively about the spherical or $X=0$ solution. The purpose of this section is to find
the effective coupling about these vacua and discuss under what conditions the expansion around
these vacua can be trusted.

Let us first consider the spherical vacuum. Expanding \eqref{Hlightcone} about the $X^i=Rx^i$ 
solution, we obtain the interaction terms which are from cubic up to sixth order in fluctuations 
$Y^i$ or $X^a$. In order to read the coupling constant, however, we should redefine (rescale) the 
fluctuations so that the quadratic part of the Hamiltonian takes the standard canonically 
normalized form of $\sum_l \hbar \omega_l a_l^\dagger a_l$, where $a^\dagger_l$ is the 
corresponding creation operator and $\omega_l$ is the mass of the mode, which in our case is 
$\mu(l+1)$. For this we need to rescale $Y^i$ and $X^a$ as
\be\label{canonical}
Y^i,X^a\to \frac{1}{\sqrt{\mu p^+}} Y^i, X^a\ .
\ee
As can be seen from \eqref{potential-fermion}, for the fermions no rescaling is needed. It is 
straightforward to see that the cubic term is suppressed by a factor of $g_{eff}$, and likewise 
terms of order $n$ in fields are accompanied by a factor of $g^{n-2}_{eff}$, where
\be\label{coupling}
g_{eff}=\frac{1}{\mu p^+ {\sqrt g_s}}=\frac{1}{R\sqrt{\mu p+}}\ .
\ee
(Note that energy is measured in units of $\mu$ and hence one should take out a factor of $\mu$ 
from the potential. This can be done systematically if we scale time with $1/\mu$.) 

One might rewrite $g_{eff}$ in terms of the BMN gauge theory parameters, $J$, $N$ and $g^2_{YM}$,
where \cite{Sadri:2003pr}
\[
\frac{1}{(\mu p^+)^2}=\frac{g^2_{YM} N}{J^2}\equiv \lambda'\ ,\ \ \ \ 
g_2\equiv \frac{J^2}{N}\ .
\]
Then $R^2=\sqrt{\lambda'} g_2$ and $g_{eff}^2=1/g_2$. Noting that $g_2$ is the genus counting 
parameter for strings on plane-waves, \eqref{coupling} suggests that our giant graviton theory is somehow S-dual to string theory on the 
plane-wave.\footnote{
It is interesting to note that the three point function of ${\cal O}_J^{{S^5}}$ \eqref{ggsph}
is given by \cite{Takayanagi:2002nv}
\be\label{ggthreepntfunc}
\langle {{\cal O}}_J^{{S^5}}\bar{\cal O}_{rJ}^{{S^5}} \bar{{\cal O}}_{(1-r)J}^{{S^5}}\rangle\simeq
e^{-g_2 r(1-r)/2} \sim e^{\frac{-r(1-r)}{g^2_{eff}}} \, ,
\ee
where $0\leq r <1$ and by $\simeq$ we mean that the result is presented after the BMN limit,
i.e. $J,N\to\infty$ and $J^2/N={\rm fixed}$. \eqref{ggthreepntfunc} shows that the above three 
point function corresponds to tunneling between two different giant graviton states which is a 
non-perturbative (instanton) effect in the giant graviton theory.}

One may repeat the same analysis for the $X=0$ vacuum, for which we should use the 
same scaling as above and hence we again end up with the same coupling as \eqref{coupling}.
We caution that the above coupling should be thought of as a ``bare'' coupling and 
in a properly  quantized  system this coupling may be dressed with some other factors of 
$\mu p^+$ and also this dressing factor can be different for different vacua.
In this respect the situation is quite similar to the membrane case which was analyzed in detail 
in \cite{Dasgupta:2002hx}. However, in our case we do not know how to quantize the Nambu brackets.

\section{Gauge Theory on Giant Gravitons}
\label{gauge-theory-on-GGs}
 
We are now ready to include the contribution from the gauge fields on the brane.
The analysis of section \ref{Lightcone-Hamiltonian} is modified slightly in this case when 
$F_{\hmu\hnu}=(d A)_{\hmu\hnu}$
is turned on. The equations giving the metric on the brane as the pullback of the
plane-wave space-time metric are of course unaffected, but the determinant appearing
in the DBI action receives contributions from the gauge field strength. We write
\[
M_{\hmu\hnu}=G_{\hmu\hnu}+F_{\hmu\hnu} \, ,
\]
where as before, $G_{\hmu\hnu}$ is the pullback of the space-time
metric given by equation \eqref{pullback-metric}. As a part of the light-cone gauge fixing, as 
we did in section \ref{LCgauge-fixing}, we set $G_{0r}=0$ and hence
\be
M_{00}=G_{00}\ ,\ \ \ \ \
M_{0r}=F_{0r}=-F_{r0}\equiv E_r\ ,\ \ \ \
M_{rs}=G_{rs}+F_{rs} \ ,
\ee
where $E_r$ is the electric field and $G_{00}$ is still given by \eqref{dtauX-}.
The Chern-Simons terms and also the fermionic contributions \eqref{potential-fermion} are 
unaffected by the appearance of the gauge fields.
The contribution of the gauge field to the momentum conjugate to $X^-$
results in a modification of \eqref{pplus}, as
\be \label{pplus-again}
  p^+ = - \frac{1}{g_s} M^{00} \sqrt{- \: det \: M} \, ,
\ee
but now $M^{00} \ne 1/M_{00}$ because of the off-diagonal electric field appearing in
$M_{\hmu\hnu}$; it becomes
\be\label{M00}
  M^{00} = \frac{det \: (G_{rs}+F_{rs})}{det \: M}\ .
\ee
The determinant appearing in the action is also modified
\be
  det \: M = \Bigl(det \: (G_{rs}+F_{rs})\Bigr)\left( G_{00} + E_r G^{rs} E_s \right) .
\ee
Using \eqref{pplus-again}, we can write
\be \label{detgrs}
  - \frac{det \: (G_{rs}+F_{rs})}{(p^+ g_s)^2} =
  G_{00} + E_r G^{rs} E_s \, .
\ee
The expression for the momentum conjugate to $X^+$ (the light-cone Hamiltonian)
remains as in \eqref{Hlc}, but with $p^+$ now given by \eqref{pplus-again}, i.e. 
\[ 
  P^- = p^+ \left( \partial_\tau X^- + \mu^2 X^I X^I \right)-
\frac{1}{6}\epsilon^{rps}C_{+IJK}\partial_r X^I\partial_p X^J\partial_s X^K  .
\]
The momenta conjugate to $X^I$ \eqref{PI} are unaffected.
We can solve for $\partial X^-$ in terms of $X^I$ and their conjugate momenta $P_I$, and also $E_r$ and 
$F_{rs}$, using \eqref{pplus-again}, \eqref{M00} and \eqref{detgrs}. Since the 
computations are very similar to those of section \ref{LCgauge-fixing} we do not repeat them here.
Gathering all the terms, the total light-cone Hamiltonian density becomes
\be\label{full-LC-Hamiltonian}
\begin{split}
  \mathcal{H}_{l.c} &= \frac{P_I^2}{2 p^+} + \frac{1}{2} \mu^2 p^+ X_I^2 +
  \frac{1}{2 p^+ g_s^2} det \: (G_{rs} +F_{rs})+
  \frac{1}{2} p^+ E_r G^{rs} E_s\\
& +
\frac{\mu}{6g_s} \Bigl(
  \epsilon^{i j k l} X^i\{ X^j, X^k, X^l \}+ \epsilon^{a b c d} X^a\{ X^b, X^c, X^d \} \Big)
  .
\end{split}
\ee

\subsection{Spectrum of small fluctuations of the gauge field}\label{photon-spectrum-section}

{}From \eqref{full-LC-Hamiltonian} it is readily seen that $X=0$ and $X^i=Rx^i$ (and of 
course together with $E_r=F_{rs}=0$) are still the only zero energy configurations. Then one may 
expand the theory about each of these vacua. The spectrum of $X^i$, $X^a$ modes is the same 
as those we studied in section \ref{bosonic-spectrum}. (This statement is also true for fermionic 
modes. In fact one can show that the full supersymmetric version of the Hamiltonian 
\eqref{full-LC-Hamiltonian} is obtained by adding \eqref{potential-fermion} to
\eqref{full-LC-Hamiltonian}. This in particular means that fermions do not directly couple to 
gauge fields. The latter could be understood by noting that we are only dealing with a $U(1)$ 
gauge theory where fermions sit in the adjoint representation, i.e. they are neutral.)

For the $X=0$ vacuum, there are no gauge field contributions, because the gauge field terms
only appear in quartic or higher powers. For example the induced metric $G_{rs}$ is 
second order in $X$ fluctuations and hence the $E_r G^{rs} E_s$ term is at least quartic.
This is compatible with our earlier discussions in section \ref{fermionic-modes}, that the 
fluctuations of $X^i$'s, $X^a$'s and the corresponding fermionic modes, complete a \super (short) 
multiplet. In this case, gauge fields only couple to ``scalar'' bosonic modes with the 
``bare'' coupling given by \eqref{coupling}.

As for the $X^i=Rx^i$ vacuum, expanding \eqref{full-LC-Hamiltonian} up to second order in all 
fluctuations we obtain
\be\label{calH(2)}
{\cal H}^{(2)}_{l.c.}= {H}^{(2)}_{l.c.}+ \frac{1}{2\mu g_s}\left(E_r g^{rs}_0 E_s+ 
\mu^{2}B_r g^{rs}_0 B_s\right) \, ,
\ee
where  ${H}^{(2)}_{l.c.}$ is given in \eqref{Hlightcone},  $g^{rs}_0$ is the (inverse) metric on 
the unit three-sphere and $g_0^{rp}B_p=\epsilon^{rqs}F_{qs}$ is the magnetic field.
It is worth noting that ${\cal H}^{(2)}_{l.c.}$ (plus \eqref{potential-fermion}) is exactly the 
Hamiltonian for an ${\cal N}=4$ $U(1)$ gauge theory on $\mathbb{R} \times S^3$
(the latter action may be found in \cite{Blau:2000xg}). 
Among the fields in the 
four dimensional ${\cal N}=4$ gauge multiplet the gauge field, four scalars, the $X^a$ modes, and 8 fermions (which are in the correct representation, e.g. see \cite{Sadri:2003pr}) are explicit. The other two scalar modes, however, are a combination of $Y^i$'s. Although the explicit 
expressions defining these two scalars in terms of $Y^i$'s is not so simple, as we 
discussed in section \ref{X=R-spectrum}, $Y^i$'s only contain two physical modes, with the masses 
equal to the other four scalar, $X^a$ modes, and also the fermions. Moreover, the above argument
would imply that the $SO(4)$ symmetry rotating $X^a$'s among each other, can be generalized to 
$SO(6)$ including these other two scalar modes, giving rise to the full $R$-symmetry group of the
${\cal N}=4$ gauge theory.

In the same manner that we argued that the fluctuation modes of section \ref{X=R-spectrum} 
complete a \super multiplet, we should also work out the spectrum of the gauge fields, i.e. 
photons, on the three-sphere, and show that the two polarizations of the photon have the same 
``mass'' as the fermions and scalars. Let us start with the equation of motion for the gauge 
field, $A_{\hmu}$:
\[
\left(g^0_{\hmu\hnu}\nabla^2-\frac{1}{2}
(\nabla_{\hmu}\nabla_{\hnu}+ \nabla_{\hnu}\nabla_{\hmu})\right)A^{\hnu}=0 \ .
\]
We choose to work in Coulomb gauge, i.e. $A_0=0,\ \nabla_{\hmu}A^{\hmu}=0$.\footnote{
We would like to point out that working with the light-cone 
gauge in the bulk, as we have done here, {\it does not} necessarily imply that for the  
worldvolume gauge theory we have also fixed the same gauge. In fact these are two independent
gauge symmetries and the $U(1)$ gauge theory should be fixed separately.} 
Next we note that $\nabla_{r}$ and the gradient on the unit three-sphere do not commute. In fact
they commute to the Riemann tensor. For a unit sphere
$R_{rs}=2 \: g_{rs}$, where $R_{rs}$ is the Ricci tensor,
and we have
\be\label{guage-field-eom}
\left( \omega^2- \mu^2(\nabla_r^2 +1)\right) A_s=0 \, .
\ee
If we take $A_r$ to be in the spin $l$ representation of $SO(4)$, i.e. $\nabla^2 
A^l_r=l(l+2)A^l_r$, the spectrum of the two photon polarizations is obtained to be
\be\label{photon-spectrum}
\omega=\mu (l+1),\ \ \quad l\geq 1.
\ee
As we see  the ``mass'' for photons has a purely geometric origin. (The same is also true for 
fermions). This is in contrast with that of scalars, where we have an explicit mass term.
Equation \eqref{photon-spectrum}, as we expected from the superalgebra arguments, leads exactly 
to the same spectrum as scalars and fermions. These two photon modes together with the other 
excitations studied in section \ref{X=R-spectrum} complete a multiplet of the
\super algebra.

\section{BIGGons: BIons on Giant Gravitons}\label{BIGon}
\label{biggons}

In this section we focus on a $U(1)$ gauge theory on $\mathbb{R} \times S^3$
and study static, BPS charge configurations. We cannot have non-zero electric net charge ($S^3$ is 
compact), however, higher-pole configurations are allowed. We study the dipole configurations in 
some detail. Looking for BPS configurations, we are forced to turn on the scalar fields as well. 
This is a direct generalization of Callan-Maldacena \cite{Callan:1997kz} and Gibbons' BIons 
\cite{Gibbons:1997xz} argument to the ``compact'' branes.
These configurations, from the point of view of an observer far away in the bulk, have the 
interpretation of (fundamental) strings piercing the $S^3$. The points of attachment (the north and south pole of the three-sphere), carry positive and negative 
electric charge.
The open strings, locally, can be thought of as Polchinski's open strings ending on the brane 
(giant graviton in our case) with Dirichlet boundary conditions \cite{Polch}.
As evidence that this configuration is really a string we show that the 
energy is proportional to the distance to the brane (at least for far 
distances).

\subsection{Solutions}
\label{solutions}

We would now like to consider placing charges on the giant graviton. 
We take for the embedding coordinates  of a unit three-sphere
\be \label{coords}
\begin{split}
  x_1 &=   \sin \psi \: \sin \theta \: \cos \phi \\
  x_2 &=   \sin \psi \: \sin \theta \: \sin \phi \\
  x_3 &=  \sin \psi \: \cos \theta \\
  x_4 &=  \cos \psi
\end{split}
\ee
with $0 \le \psi, \theta \le \pi$ and
$0 \le \phi \le 2 \pi$. The metric on the three-sphere in the coordinate system \eqref{coords} is
\be \label{metric}
  ds^2 \: = \: d \psi^2 + \sin^2 \psi   d \Omega_2^2 \, ,
\ee
with $d \Omega_2^2=\left( d \theta^2 + \sin^2 \theta \: d \phi^2 \right)$ the metric
on the unit two-sphere, and $\sqrt{\det g}=\sin^2 \psi \: \sin \theta$. 
The Laplacian acting on a scalar field $\Phi$ 
is $\nabla^2 \Phi=\frac{1}{\sqrt{\det g}} \partial_\mu 
\left( \sqrt{\det g}
g^{\mu \nu} \partial_\nu \Phi \right)$, which on the three-sphere for $\Phi$'s with only $\psi$ 
dependence becomes
\be
  \nabla^2 \Phi (\psi) = \frac{1}{\sin^2 \psi}
  \partial_\psi \left( \sin^2 \psi \: \partial_\psi \Phi(\psi) \right) \, ,
\ee

The three-sphere is compact, and hence the giant graviton cannot support single
charges. It does, however, support dipoles (as well as higher poles) with vanishing total charge. 
Consider a dipole with two opposite charges placed at the two poles ($\psi=0,\pi$).
The charge density of such a configuration, as seen by the gauge theory, is
\be \label{sources}
  \rho \: = \: \frac{Q}{\sin^2 \psi}
  \Big[ \delta (\psi) - \delta (\psi - \pi) \Big]
  \delta (\cos \theta) \delta (\phi) \, .
\ee

A few words about the normalizations of the fields are in order:
The scalar and gauge fields appearing in the Hamiltonian \eqref{quad:Hamiltonian} and 
\eqref{calH(2)}
are not canonically normalized. The normalization of the gauge field in
\eqref{calH(2)} is such that the gauge theory action carries an overall factor of
$1/g_{YM}^2$, where $g_{YM}=\mu \sqrt{g_s}$,
a convenient choice for studying gauge theories on curved backgrounds.
Canonical normalization of the gauge field can be achieved by taking
$A_\mu \rightarrow \mu \sqrt{g_s} A_\mu$.
The coupling of the gauge field to the charges in the gauge theory is of the form
$J \cdot A$, with $J_0$ the charge density $\rho$, which carries the same units
as the charge $Q$ since the angular coordinates are dimensionless.
The scalar fields in \eqref{quad:Hamiltonian}, as stated in \eqref{canonical}, can be normalized 
canonically by taking $X \rightarrow R \: g_{eff} \: \Phi$.
This is the normalization in which the scalar field couples to the sources in the
same way as the gauge field, via \eqref{sources}.
Noting that the energy is measured in units of $\mu$ (a choice of scale for the
time coordinate), we find that the charges, as seen from the DBI action
\eqref{DBI-action}, are measured in units of $\mu \sqrt{g_s}$, so $q=Q \mu \sqrt{g_s}$, with $Q$ 
dimensionless.
It is the canonical fields (scalar and gauge) that are sourced by $Q$.
The fact that the gauge and scalar fields enter the Hamiltonian with different scales
is an artifact of the choice of relative normalizations of the fields in the DBI
action \eqref{DBI-action}.
In fact, the normalizations in the DBI action are chosen to
reproduce the correct field normalizations in the gauge theory for the case
of a flat background, but in a curved background do not reproduce the standard
normalizations for a gauge theory coupled to a fixed curved metric, as is the case for
the SYM theory on $\mathbb{R} \times S^3$.
We also remind the reader that $\alpha^\prime \mu p^+ g_s$ is a dimensionless quantity,
so in our units, where we have set $\alpha^\prime = 1/2 \pi$, $\mu p^+$ and $g_{eff}$ are
both dimensionless. 
{}From this point on we shall deal only with canonically normalized fields.

We study the electrostatic problem for the gauge field in a gauge where $A_0=\Lambda$.
The equation of motion for the gauge field, arising from the quadratic
Hamiltonian \eqref{calH(2)}, is simply
Poisson's equation, which in appropriate units requires $\nabla^2 \Lambda = - \rho$.
We take as our ansatz a field $\Lambda(\psi)$ which is a constant along the
two angular directions $\theta,\phi$.
The solution is
\be \label{gauge:soln}
  \Lambda (\psi) = Q \cot \psi  .
\ee
As in the discussion of \cite{Callan:1997kz}, exciting the gauge field alone
would not result in a BPS configuration (supersymmetry implies a relation
between the profile of the gauge field and the other fields in the
$\mathcal{N}=4$ supermultiplet). The solution can be made BPS by turning
on a non-trivial profile for the scalars, keeping a constant vanishing
background fermion field.

To find the scalar profile, it will prove useful to consider the quadratic part of the 
Hamiltonian, expanded around the spherical vacuum, restricted to field configurations which depend 
only on the radial direction. 
The potential for the scalar field describing the radial profile can then be written
\be \label{scalar:pot}
  V_{\Phi} = \frac{1}{2}
  \left(
  \Phi^2 + (\nabla_{S^3} \Phi)^2
  \right) \, ,
\ee
where the square is with respect to the metric on the
three-sphere of unit radius.
The spikes can't go off to infinity in just any direction because of the
potential from the plane-wave, but they can extend off to infinity along
$X^-$, which can be taken as the radial direction.

The equation of motion for the field $\Phi$ in the gauge theory, in the
presence of a dipole charge configuration is simply
\be \label{scalar:eom}
  \nabla_{S^3}^2 \Phi - \Phi = \frac{Q}{\sin^2\psi}   \Big[ \delta (\psi) + \delta (\psi - \pi) 
\Big]
  \delta (\cos \theta) \delta (\phi)  .
\ee
The right hand side represents the sourcing of the scalar field $\Phi$ by the charges,
as required by supersymmetry. The fact that the two sources contribute with the
same sign can be seen (as we will show below) from requiring that the solutions
of these equations, together with the electric field, form a BPS configuration.

\begin{figure}[ht]
\centering
\epsfig{figure=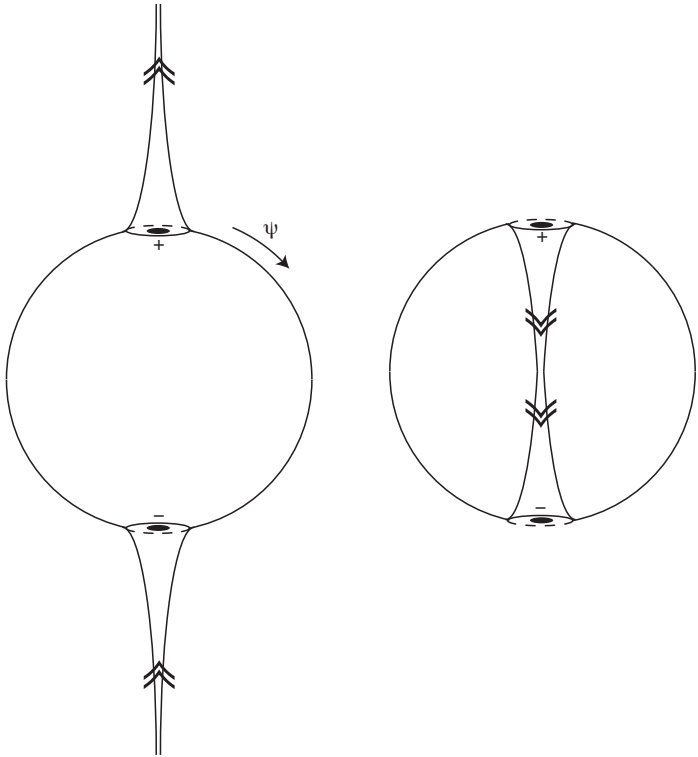,width=92mm,height=100mm}
\begin{center}
\caption{Two dipole configurations. The left one has an interpretation of two fundamental strings 
extending off the giant graviton to infinity. The configuration on the right is unstable and
has no such string interpretation. The sign of the charges are indicated and the arrows
denote the direction of flux of the electric field.}
\label{fig-dipoles}
\end{center}
\end{figure}

The equation of motion for the scalar field \eqref{scalar:eom} is solved by taking
(see the Appendix for more details)
\be \label{scalar:soln}
  \Phi (\psi) = \frac{Q}{\sin \psi} \, .
\ee
Measured with respect to the origin in spherical coordinates, the profile of the
spike is given by $R\pm \Phi$, with $R$ the radius of the giant graviton
(see Figure \ref{fig-dipoles}).
We would like to note that in our case the profile of $\Phi$ and the Coulomb 
potential $\Lambda$, \eqref{scalar:soln} and \eqref{gauge:soln} are different; this should be 
contrasted with the usual BIon case \cite{Callan:1997kz}. 
Near the north pole (with $\psi \approx 0$), the solution for the scalar field and gauge field
are similar, while at the south pole ($\psi \approx \pi$), they differ by a sign, and for $\psi$ 
away from the poles, each solution interpolates smoothly between the solutions in the two regions. 
The scalar field is blind to the sign of the charges which source it, while the
gauge field is not. 

\begin{figure}[ht]
\centering
\epsfig{figure=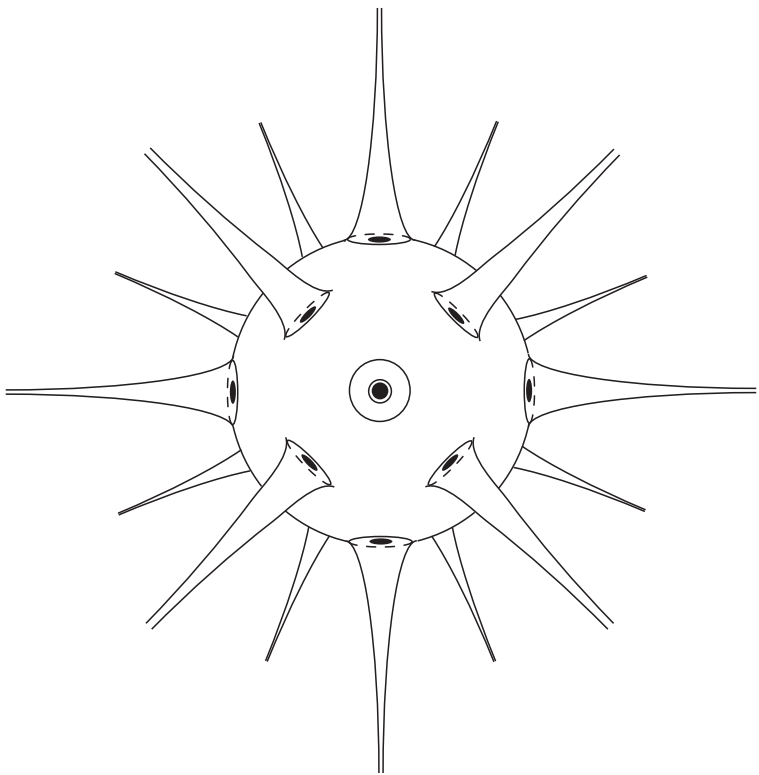,width=100mm,height=102mm}
\begin{center}
\caption{A generic giant graviton with multiple spikes, suggesting the
hedge-hog title.}
\label{fig-hedge-hog}
\end{center}
\end{figure}

The size of the throat, as seen by the DBI action, is $R \: g_{eff}$, the coupling
for the canonically normalized fields. Explicitly, the solution for the radial direction $X$ is
\be\label{Xspike}
X=R\left(1\pm Q\frac{g_{eff}}{\sin\psi}\right).
\ee
Interestingly, the corrections to the shape of the strings arising from
non-linearities are suppressed relative to the tree-level shape by the same
scale that sets the throat size, i.e. $g_{eff}$. This basically means that the perturbative 
expansion of the light-cone Hamiltonian (up to quadratic order) is a good one as long as
the size of the throat is much smaller than the giant graviton itself.
The non-linearities and interaction terms can in fact modify the form of the potential around a 
given vacuum such that one solution, explicitly the solution corresponding to the choice of 
minus sign in \eqref{Xspike}, is destabilized, as 
happens to the dipole solution (diagram on right of Figure \ref{fig-dipoles}),
where the strings enter the interior and meet. The energy of this arrangement can be
lowered by moving the end-points of the strings closer to each other, and the
charges at the ends would eventually annihilate, leaving behind a giant graviton
with no strings attached. For the dipole where the strings extend off to infinity, i.e. the 
solution with plus sign in \eqref{Xspike},
the layout of the strings at opposite ends (the diagram on the left of Figure
\ref{fig-dipoles}) is in fact a minimum, and remains so even when the interactions
are included, with the interactions only modifying the profile of the string
at the junction.
\begin{figure}[ht]
\centering
\epsfig{figure=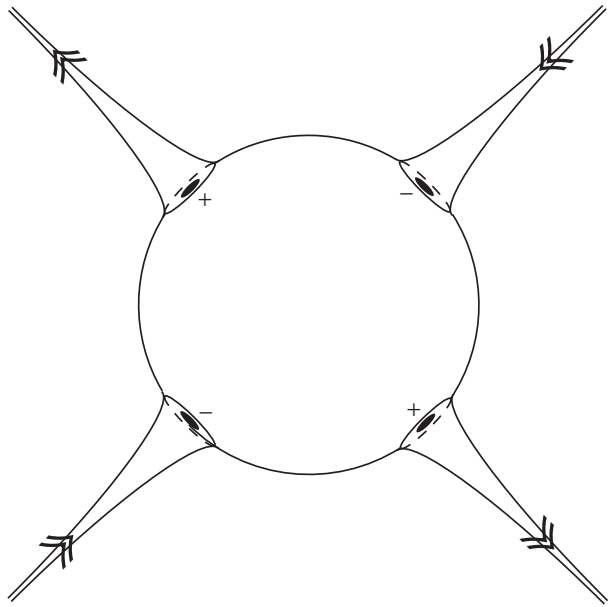,width=80mm,height=80mm}
\begin{center}
\caption{A quadrapole configuration of charges and the associated fundamental strings on the
giant graviton.}
\label{fig-quadrapole}
\end{center}
\end{figure}
Higher pole solutions can be analogously constructed  (see Figure 
\ref{fig-quadrapole} for a quadrapole, and Figure \ref{fig-hedge-hog} for a more general
``hedge-hog'' configuration).
When the coupling is of order the separation of the strings attached to the giant graviton,
the end-points can meet and the string can separate from the giant graviton.
{}From the gauge theory point of view, this corresponds to the charges
annihilating each other.
For finite coupling (and hence finite throat size), there will be a maximum number of
strings which can attach to the giant graviton, and hence a maximum pole configuration
in the gauge theory. The smallest size which can be effectively probed by the
open strings is set by $g_{eff}$, and this leads to fuzziness of the giant gravitons in view of 
an open string probe.

One may also consider configurations which are sourced by magnetic dipoles 
(and higher poles). Such configurations correspond to S-dual solutions to the ones
considered above, where the strings ending on the giant graviton are D-strings.
Dyonic configurations with both electric and magnetic sources can also be envisioned.
More general configurations, where several giant gravitons are coincident,
can also be constructed, and by analogy to the general case of coincident
D-branes, would give rise a non-Abelian gauge theory on their worldvolume.

\subsection{Energy}

We would now like to give an interpretation to the spikes we found as solutions
of the quadratic Hamiltonian in section \ref{solutions}.
To do so we consider the energy of such a configuration.
To find the energy density, we use the solutions \eqref{gauge:soln} and \eqref{scalar:soln} for the 
gauge field configuration and radial profile of the giant
graviton in the Hamiltonian density, then integrate the
density to find the total energy.
For the dipole, there are two solutions, one for which the spike extends
off to infinity away from the giant graviton, and one where the spikes enter the interior and join 
(see Figure \ref{fig-dipoles}).
We consider both configurations.
The resultant energy ${\cal E}$, in units of $\mu$, for the first configuration is
\be
  {\cal E} = 4 \pi Q^2 \cot \epsilon \, ,
\ee
and the scalar and gauge fields contribute equally to the energy.
We have integrated along the $\psi$ direction from $\epsilon$ to $\pi/2$. The $\epsilon$
serves as a cut-off, since the total energy would diverge; we are interested in the scaling
of this energy with length as the cut-off is removed.
The $\pi/2$ captures one string (the other string would give an equal contribution).
For small $\epsilon$, the result scales as
\be
  {\cal E} \sim \Phi(\epsilon)
\ee
up to some fixed numerical coefficients.
In other words, the energy per unit length is the same as the tension of the fundamental
string.\footnote{In our units, the string tension $T \sim 1$.}
We expect also that the spectrum of small fluctuations for the string should
reproduce the spectrum of massive modes of the open fundamental string in the plane-wave 
background \cite{Balasubramanian:2002sa}. 
\begin{figure}[ht]
\centering
\epsfig{figure=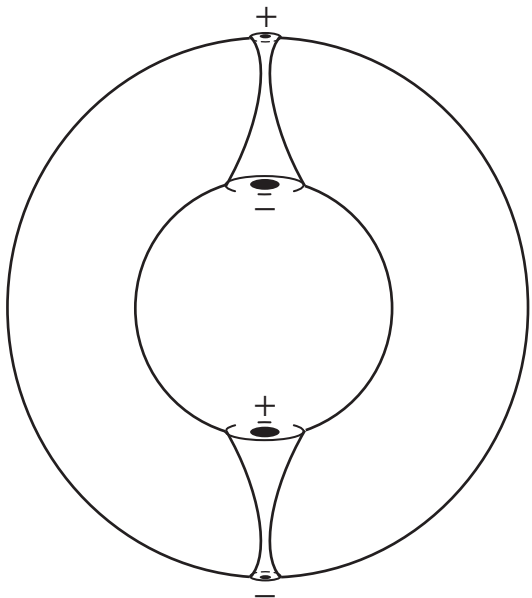,width=70mm,height=83mm}
\begin{center}
\caption{Two giant gravitons of different radii, connected by ``strings''. In the limit
that the radii are equal and the giant gravitons become coincident, the gauge
symmetry of the worldvolume theory is enhanced.}
\label{fig-double}
\end{center}
\end{figure}

The profile where the ``strings'' enter into the interior of the giant graviton is given by
$R - \Phi$, with $\Phi$ the same as for the outgoing strings, but the range of $\psi$ is now
restricted such that $R - \Phi$ is limited to only one ``hemisphere'' inside
the giant graviton. At the lower cutoff for $\psi$, the string joins onto the other
string originating at the other charge. In other words, $\sin^{-1}(g_{eff})\leq \psi\leq
\pi-\sin^{-1}(g_{eff})$ and only when the size of the throat is very small compared to the size of 
the giant graviton
(i.e. when $g_{eff} \ll 1 $) this spike can be interpreted as a fundamental
string. For any finite value of $g_{eff}$, the profile never reaches that of a
fundamental string.

In any case, as we have already discussed, this configuration is unstable. The
dipole where the strings run off to infinity is stabilized by the fact that the endpoints of the
strings have their boundary conditions fixed at infinity, and any small perturbation
of their junction increases the length of the string and hence the energy. 

Given a total light-cone momentum $p^+$, one may distribute it among some number of giant 
gravitons, that is a configuration of concentric giant gravitons. One may wonder whether 
in the limit when two of these giant gravitons become coincident, analogously to the case of 
D-branes, one should expect enhancement of the $U(1)$ gauge symmetry to $U(2)$. That is possible if the unstable spikes (strings) depicted in Figure \ref{fig-double} become massless in the 
coincident limit, a fact which is confirmed by our energy analysis. 
The decay rate of these spikes depends on $g_{eff}$ as well as the difference in the radii of the two giant gravitons. Based on energy arguments we expect it to be proportional to 
$\sqrt{g_s}$ as well as the difference of inverse radii squared of the two giant gravitons.

\subsection{Supersymmetry}

The spin connection one-forms on the three-sphere can be deduced from the metric
\eqref{metric}
\be
  \Omega^{12} = - \cos \psi d \theta \, , \ \ \ \ \
  \Omega^{23} = - \cos \theta d \phi \, , \ \ \ \ \
  \Omega^{31} = \cos \psi \sin \theta d \phi \, ,
\ee
and the supersymmetry variation of the gaugino for the abelian theory is
\cite{Okuyama:2002zn,Nicolai:1988ek}
\be \label{susy-variation}
  \delta_\epsilon \lambda =
  \Big(
  \frac{1}{2} F_{\mu \nu} \Gamma^{\mu \nu} -
  (\partial_\mu \Phi^m) \Gamma^m \Gamma^\mu -
  \frac{1}{2} \Phi_m \Gamma^m \Gamma^\mu \nabla_\mu
  \Big) \epsilon \, .
\ee
The derivatives on the fields are not gauge covariant since these fields
transform in the adjoint of $U(1)$, and hence are neutral.
Also, $\nabla_\mu = \partial_\mu + \frac{1}{4} \Omega^{ab}_\mu \Gamma_{ab}$, is the
covariant derivative with the spin connection $\Omega^{ab}_\mu$.\footnote{
Note that the indices $a,b$ are with respect to the orthonormal tangent frame, while
$\mu,\nu$ are curved indices on the worldvolume and $m$ ranges over the six
$SO(6)$ components.}
Solutions $\epsilon$ in a given background, for which this variation vanishes,
are the Killing spinors, and the number of such solutions gives the amount of supersymmetry
preserved by the background.
For the solutions we are considering, the deformation of the sphere is
independent of the angular directions along an $S^2$ of the $S^3$, and preserves
an $SU(2)$ symmetry of the $SO(4)$ invariant vacuum.

We  make use of the solution for the electric field in terms of the
scalar potential, where
\be
  F_{0 \psi} = -\partial_\psi \Lambda(\psi) = \frac{Q}{\sin^2 \psi} \, ,
\ee
and $\Phi^m \Gamma^m=\Phi(\psi) \Gamma^r$ with $r$ designating the radial direction in the
transverse directions, and with $\Phi=\frac{Q}{\sin \psi}$.
The Killing spinor equation for this background is \cite{Blau:2000xg}
\be \label{killing-spinor}
  \nabla_\mu \epsilon = \frac{1}{2}\Gamma^r \Gamma_\mu \epsilon \, ,
\ee
and the square of the Killing spinors are the Killing vectors.
The Killing spinor equation \eqref{killing-spinor} has the maximal number of solutions
\cite{Okuyama:2002zn}.
For these Killing spinors, the condition for supersymmetry \eqref{susy-variation}
reduces to\footnote{With our metric conventions, $(\Gamma^0)^2=-1$, with the other Dirac matrices 
squaring to one.}
\be \label{susy-spikes-out}
  \frac{Q}{\sin^2 \psi}
  \Big(
  \Gamma^0 + \tilde{\Gamma}^r
  \Big) \epsilon^\prime = 0 \, ,
\ee
with $\epsilon^\prime = \Gamma^\psi \epsilon$, and
\[
\tilde{\Gamma}^r = \cos \psi \Gamma^r + \sin \psi \Gamma^\psi
\] 
is a rotated Dirac matrix of unit norm, i.e. $(\tilde{\Gamma}^r)^2=1$. 
This implies that half the supersymmetries of the background plane-wave remain
unbroken by the presence of the D-brane, and the dipole configuration of the
giant graviton state with the two spikes piercing it is $1/2$ BPS, i.e. it preserves eight 
supercharges.

The $\tilde{\Gamma}^r$ 
matrix at $\psi=0$ is ${\Gamma}^r$ and hence the supersymmetry condition \eqref{susy-spikes-out}
is essentially that of the usual BIon \cite{Callan:1997kz} with, say a positive charge. 
At $\psi=\pi$, however, $\tilde{\Gamma}^r=-\Gamma^r$ reducing \eqref{susy-spikes-out} to a usual 
BIon with 
negative charge and the term proportional to $\Phi$ in \eqref{susy-variation} makes it possible 
to have a smooth supersymmetric  transition form a positive charge to a negative charge. 

There also exists a solution of the quadratic Hamiltonian for which
$\Phi=-\frac{Q}{\sin \psi}$, but with the gauge field configuration unchanged.
The condition for the existence of supersymmetry
for this configuration is
\be \label{susy-spikes-in}
  \frac{Q}{\sin^2 \psi}
  \Big(
  \Gamma^0 - \tilde{\Gamma}^r
  \Big) \epsilon^\prime = 0 \, ,
\ee
which differs from \eqref{susy-spikes-out} by the relative sign between the time-like and
radial Dirac matrices, but preserves precisely the same amount of supersymmetry as
the original configuration, and is also $1/2$ BPS. One should note that being BPS does not 
necessarily guarantee the stability of the solution, particularly when the interaction terms in the 
light-cone Hamiltonian are taken into account.\footnote{In principle the statement that BPS 
configurations are protected  should be taken with a grain of salt. It is possible that some 
multiplets which are BPS  at a given value of coupling combine into a long (ordinary non-BPS) 
multiplet and receive corrections. For explicit examples and more detailed discussion on this point 
see \cite{Dasgupta:2002ru}.}

\section{Outlook and future directions}
\label{outlook}

In this paper we have analyzed some aspects of the worldvolume theory of giant gravitons on the 
plane-wave background. Working out the spectrum of small fluctuations of the giant 
three-sphere, we argued that they fall into (short) multiplets of the \super algebra. One of the 
interesting features of the three-brane light-cone Hamiltonian \eqref{Hlightcone} is the natural 
appearance of Nambu brackets \eqref{Nambubracket}. In this point of view ``quantization'' of Nambu 
bracket \eqref{Nambubracket} would provide us with a natural quantization of the theory living on 
the giant graviton. In the membrane case the corresponding Nambu bracket is essentially a Poisson 
bracket and its quantization is possible by replacing the bracket with Matrix commutators   
\cite{Taylor:2001vb}. This quantization of Possion brackets from the membrane point of view can be 
regarded as discretization of the worldvolume, which also leads to the ``noncommutative'' 
(non-Abelian) structure of the BFSS matrix model. In the same trend quantization of the 
three-sphere giant graviton theory may provide us with an answer to the puzzle of finding a
holographic description of type IIB string theory on the plane-wave background, which supposedly 
is a Matrix theory \cite{progress} (for a summary of discussions on the matter see section IX of 
\cite{Sadri:2003pr}). 

As another aspect of the gauge theory living on giant gravitons, we studied static configurations 
which source the gauge fields and also the scalar fields. The basic building blocks of such 
objects are dipole configurations with the largest possible dipole moment being proportional to 
the size of the giant graviton. We argued that the BPS dipole configurations, from the bulk 
viewpoint, can be understood as open strings ending on the giant graviton. These are open strings 
with their two ends on the north and south pole of the three-sphere. It is evident from our 
construction that these open strings satisfy Dirichlet boundary conditions in the directions 
transverse to the brane, a natural expectation generalizing Polchinski's D-brane picture 
\cite{Polch}.
We also argued that it is possible to have dipole configurations with the spike going inside the 
three-sphere. These states are responsible for enhancing the gauge 
symmetry when two concentric giant gravitons become coincident. 

As we discussed, since at finite $g_{eff}$ the size of the throat of the spikes is finite, one would physically expect to have an upper limit on the highest multiple moment. In other words
there is a minimum area which can be probed using these open strings and also there is a minimum 
size dipole moment. This suggests that the fuzzy three sphere \cite{Ramgoolam:2001zx} is the right description of the quantized giant graviton \cite{progress}.
A description of multiple coincident giant gravitons in terms of a non-commutative
three-sphere defined as a Hopf fibration over a fuzzy two-sphere is given in
\cite{Janssen:2003ri}.

As a direct generalization of our giant hedge-hog configurations one can consider
circular D-strings in the $AdS_3\times S^3$ background (or the corresponding Penrose limit 
\cite{Russo}). In that case, however, we expect that similar to the flat D-string case 
\cite{Dasgupta:1997pu},
the spike touches the giant circle to form a ``string junction''. This leads to 
a pair of three string junctions, two of the legs of each are connected and make a deformed half circle.
This construction can then be generalized to junctions of $(p,q)$ strings and string networks 
\cite{Sen:1997xi} in  plane-wave backgrounds \cite{junc-in-progress}.

{\large{\bf Acknowledgements}}

We would like to thank Keshav Dasgupta, Michal Fabinger, Sergey Prokushkin,
and especially Leonard Susskind for fruitful discussions.
The work of M. M. Sh-J. is supported in part by NSF grant
PHY-9870115 and in part by funds from the Stanford Institute for Theoretical 
Physics. The work of D. S. is supported by the Department of Energy,
Contract DE-AC03-76SF00515.
 \renewcommand{\theequation}{A.\arabic{equation}}
  \setcounter{equation}{0}  
  \section*{Appendix:\ $SO(4)$ Harmonics in terms of usual $Y_{lm}$'s}
  \label{appendix}  

The Laplacian on the three-sphere in the coordinate system we have adopted is
\be\label{S3S2} 
\nabla^2_{S^3}=\frac{1}{\sin^2\psi}\partial_{\psi}\left(\sin^2\psi 
\partial_\psi\right)+\frac{1}{\sin^2\psi}\nabla^2_{S^2} \, ,
\ee
where $\nabla^2_{S^2}=\frac{1}{\sin\theta} \: \partial_{\theta}\left(\sin\theta 
\partial_{\theta}\right)+\frac{1}{\sin\theta}\partial^2_\phi$.
One may use \eqref{S3S2} to write $SO(4)$ harmonics in terms of the $SO(3)$ $Y_{lm}$'s. 
Explicitly, let us consider the (source free) equation of motion for the Coulomb potential 
$\Lambda$:
\be\label{nabla-chi} 
\nabla^2_{S^3} \: \Lambda (\psi, \theta, \phi)=0 \, .
\ee
Separating variables as $\Lambda (\psi, \theta, \phi)=\Lambda_l(\psi) Y_{lm}(\theta,\phi)$,
\eqref{nabla-chi} can be cast in the form
\be\label{psi-l-chi}
\frac{1}{\sin^2\psi}\partial_{\psi}\left(\sin^2\psi \Lambda_l\right)-\frac{1}{\sin^2\psi} 
l(l+1)\Lambda_l=0\ .
\ee
After the change of variable $u=\cot \psi$, \eqref{psi-l-chi} takes the form
\be\label{chi-u}
(1+u^2)\Lambda_l''-l(l+1) \Lambda_l=0\ ,
\ee
where $\Lambda^\prime=\frac{d}{d u}\Lambda$. For $l=0$, \eqref{chi-u} is simply solved by 
$\Lambda_0=u=\cot \psi$ 
(the solution we have already discussed as the dipole \eqref{gauge:soln}), and for $l=1$, 
$\Lambda_1=1+u^2=1/\sin^2\psi$. For general $l$, \eqref{chi-u} can be solved using a series 
expansion for $\Lambda_l(u)$
\[
\Lambda_l(u)=\sum_{k=0}^{l+1} a_k u^k \, ,
\]
where $a_{l+1}=1$, $a_l=0$  and
\[
a_k=\frac{(k+1)(k+2)}{l(l+1)-k(k-1)}\ a_{k+2}\ ,\ \ \ \ \ \ \ 0\leq k\leq l-1\ . 
\]

Similarly, solutions to the equation for the scalar field, namely
$\nabla^2_{S^3}\Phi-\Phi=0$ can be decomposed as 
\[
\Phi=\Phi_l(\psi) Y_{lm}(\theta, \phi) \, .
\]
Taking $v=1/\sin\psi$ and $\frac{d}{dv}\Phi=\Phi'$, then
\be\label{Phi-l-v}
v^2(v^2-1) \Phi''_l +v\Phi'_l -[l(l+1) v^2+1] \Phi_l=0\ .
\ee
For the $l=0$ case, as we  discussed in \eqref{scalar:soln}, the solution is $\Phi_{l=0}=v$, and 
for 
general $l$, as in the previous case, \eqref{Phi-l-v} 
may be solved using Taylor expansion techniques, inserting
\[
\Phi_l(v)=\sum_{k=0}^{l+1} b_k v^k
\]
into the equation. It turns out that \eqref{Phi-l-v} has only solutions for {\it even} $l$ with
$b_0=0$, $b_{l+1}=1$, and 
\[
b_k=-\frac{(k+1)^2}{l(l+1)-k(k-1)}\ b_{k+2}\ ,\ \ \ \ \ \ \ 1\leq k\leq l-1\ . 
\]
The fact that \eqref{Phi-l-v} has (polynomial) solutions only for even $l$ can physically 
be understood by noting that the source term for the scalars
is a {\it sum} of delta-functions (whereas that of the Coulomb potential is an alternating
sum, so that the total net charge is zero).

Finally, we would like to mention that in our expansions the $2^{l+1}$-poles of $SO(4)$ are 
related to $Y_{lm}$ (i.e. $2^l$-pole of $SO(3)$). For example, our ``dipole'' configurations 
correspond to the $l=0$ case. Also note that the dipole configuration can be thought of as a Dirac 
string on the sphere where $\psi=0$ corresponds to the monopole and $\psi=\pi$ corresponds to the 
end of the Dirac string tail, which in the flat space language is at infinity.



\begin{thebibliography}{99}
\bibitem{McGreevy:2000cw}
J.~McGreevy, L.~Susskind and N.~Toumbas,
``Invasion of the giant gravitons from Anti-de Sitter space,''
JHEP {\bf 0006}, 008 (2000)
[arXiv:hep-th/0003075].

\bibitem{Grisaru:2000zn}
M.~T.~Grisaru, R.~C.~Myers and O.~Tafjord,
``SUSY and Goliath,''
JHEP {\bf 0008}, 040 (2000)
[arXiv:hep-th/0008015].

\bibitem{Hashimoto:2000zp}
A.~Hashimoto, S.~Hirano and N.~Itzhaki,
``Large branes in AdS and their field theory dual,''
JHEP {\bf 0008}, 051 (2000)
[arXiv:hep-th/0008016].

\bibitem{Das:2000st}
S.~R.~Das, A.~Jevicki and S.~D.~Mathur,
``Vibration modes of giant gravitons,''
Phys.\ Rev.\ D {\bf 63}, 024013 (2001)
[arXiv:hep-th/0009019].

\bibitem{Myers:2001aq}
R.~C.~Myers and O.~Tafjord,
``Superstars and giant gravitons,''
JHEP {\bf 0111}, 009 (2001)
[arXiv:hep-th/0109127].

\bibitem{Leblond:2001gn}
F.~Leblond, R.~C.~Myers and D.~C.~Page,
JHEP {\bf 0201}, 026 (2002)
[arXiv:hep-th/0111178].

\bibitem{Balasubramanian:2001nh}
V.~Balasubramanian, M.~Berkooz, A.~Naqvi and M.~J.~Strassler,
``Giant gravitons in conformal field theory,''
JHEP {\bf 0204}, 034 (2002)
[arXiv:hep-th/0107119].

\bibitem{Corley:2001zk}
S.~Corley, A.~Jevicki and S.~Ramgoolam,
``Exact correlators of giant gravitons from dual N = 4 SYM theory,''
Adv.\ Theor.\ Math.\ Phys.\  {\bf 5}, 809 (2002)
[arXiv:hep-th/0111222].


\bibitem{Balasubramanian:2002sa}
V.~Balasubramanian, M.~x.~Huang, T.~S.~Levi and A.~Naqvi,
``Open strings from N = 4 super Yang-Mills,''
JHEP {\bf 0208}, 037 (2002)
[arXiv:hep-th/0204196].


\bibitem{Ouyang:2002vg}
P.~Ouyang,
``Semiclassical quantization of giant gravitons,''
arXiv:hep-th/0212228.



\bibitem{Berenstein:2003ah}
D.~Berenstein,
``Shape and holography: Studies of dual operators to giant gravitons,''
Nucl.\ Phys.\ B {\bf 675}, 179 (2003)
[arXiv:hep-th/0306090].



\bibitem{Blau:2001ne}
M.~Blau, J.~Figueroa-O'Farrill, C.~Hull and G.~Papadopoulos,
``A new maximally supersymmetric background of IIB superstring theory,''
JHEP {\bf 0201}, 047 (2002)
[arXiv:hep-th/0110242].

M.~Blau, J.~Figueroa-O'Farrill, C.~Hull and G.~Papadopoulos,
``Penrose limits and maximal supersymmetry,''
{\it Class. Quant. Grav.} {\bf 19} (2002) L87, hep-th/0201081;

M.~Blau, J.~Figueroa-O'Farrill and G.~Papadopoulos,
``Penrose limits, supergravity and brane dynamics,''
{\it Class.\ Quant.\ Grav.}  {\bf 19}, 4753 (2002), hep-th/0202111.



\bibitem{BMN}
D. Berenstein, J. Maldacena, H. Nastase, ``Strings in flat space and pp 
waves from ${\cal N}=4$ Super Yang Mills,'' {\it JHEP} {\bf 0204} (2002) 
013, hep-th/0202021. 


\bibitem{Metsaev:2001bj}
R.~R.~Metsaev,
``Type IIB Green-Schwarz superstring in plane wave Ramond-Ramond  background,''
Nucl.\ Phys.\ B {\bf 625}, 70 (2002)
[arXiv:hep-th/0112044].


\bibitem{Metsaev:2002re}
R.~R.~Metsaev and A.~A.~Tseytlin,
``Exactly solvable model of superstring in plane wave Ramond-Ramond  background,''
Phys.\ Rev.\ D {\bf 65}, 126004 (2002)
[arXiv:hep-th/0202109].

\bibitem{Sadri:2003pr}
D.~Sadri and M.~M.~Sheikh-Jabbari,
``The plane-wave / super Yang-Mills duality,''
arXiv:hep-th/0310119.



\bibitem{Takayanagi:2002nv}
H.~Takayanagi and T.~Takayanagi,
``Notes on giant gravitons on pp-waves,''
JHEP {\bf 0212}, 018 (2002)
[arXiv:hep-th/0209160].


\bibitem{Callan:1997kz}
C.~G.~.~Callan and J.~M.~Maldacena,
``Brane dynamics from the Born-Infeld action,''
Nucl.\ Phys.\ B {\bf 513}, 198 (1998)
[arXiv:hep-th/9708147].

\bibitem{Gibbons:1997xz}
G.~W.~Gibbons,
``Born-Infeld particles and Dirichlet p-branes,''
Nucl.\ Phys.\ B {\bf 514}, 603 (1998)
[arXiv:hep-th/9709027].


\bibitem{Book}
J. Polchinski, ``String Theory, Vol. II: Superstring theory and beyond,'' Cambridge University 
Press 1998. 










\bibitem{Maldacena:2002rb}
J.~Maldacena, M.~M.~Sheikh-Jabbari and M.~Van Raamsdonk,
``Transverse fivebranes in matrix theory,''
JHEP {\bf 0301}, 038 (2003)
[arXiv:hep-th/0211139].

\bibitem{Dasgupta:2002ru}
K.~Dasgupta, M.~M.~Sheikh-Jabbari and M.~Van Raamsdonk,
``Protected multiplets of M-theory on a plane wave,''
JHEP {\bf 0209}, 021 (2002)
[arXiv:hep-th/0207050].


\bibitem{Kallosh:1998zx}
R.~Kallosh, J.~Rahmfeld and A.~Rajaraman,
``Near horizon superspace,''
JHEP {\bf 9809}, 002 (1998)
[arXiv:hep-th/9805217].

\bibitem{Dasgupta:2002hx}
K.~Dasgupta, M.~M.~Sheikh-Jabbari and M.~Van Raamsdonk,
``Matrix perturbation theory for M-theory on a PP-wave,''
JHEP {\bf 0205}, 056 (2002)
[arXiv:hep-th/0205185].

\bibitem{Metsaev:2002sg}
R.~R.~Metsaev,
``Supersymmetric D3 brane and N = 4 SYM actions in plane wave backgrounds,''
Nucl.\ Phys.\ B {\bf 655}, 3 (2003)
[arXiv:hep-th/0211178].

\bibitem{Blau:2000xg}
M.~Blau,
``Killing spinors and SYM on curved spaces,''
JHEP {\bf 0011}, 023 (2000)
[arXiv:hep-th/0005098].


\bibitem{Polch}
Joseph Polchinski, ``Dirichlet-Branes and Ramond-Ramond Charges,''
{\it Phys. Rev. Lett.} {\bf 75} (1995) 4724, hep-th/9510017.



\bibitem{Okuyama:2002zn}
K.~Okuyama,
``N = 4 SYM on $R \times S^3$ and pp-wave,''
JHEP {\bf 0211}, 043 (2002)
[arXiv:hep-th/0207067].

\bibitem{Nicolai:1988ek}
H.~Nicolai, E.~Sezgin and Y.~Tanii,
``Conformally Invariant Supersymmetric Field Theories On $S^P \times S^1$ And Super P-Branes,''
Nucl.\ Phys.\ B {\bf 305}, 483 (1988).


\bibitem{Taylor:2001vb}
W.~Taylor,
``M(atrix) theory: Matrix quantum mechanics as a fundamental theory,''
Rev.\ Mod.\ Phys.\  {\bf 73}, 419 (2001)
[arXiv:hep-th/0101126].



\bibitem{progress}
M.M. Sheikh-Jabbari, ``Plane-wave Matrix String Theory,'' {\it Work in Progess}.

\bibitem{Ramgoolam:2001zx}
S.~Ramgoolam,
``On spherical harmonics for fuzzy spheres in diverse dimensions,''
Nucl.\ Phys.\ B {\bf 610}, 461 (2001)
[arXiv:hep-th/0105006].

\bibitem{Janssen:2003ri}
B.~Janssen, Y.~Lozano and D.~Rodriguez-Gomez,
Nucl.\ Phys.\ B {\bf 669}, 363 (2003)
[arXiv:hep-th/0303183].


\bibitem{Russo}
J.G. Russo, A.A. Tseytlin, ``On solvable models of type IIB superstring in 
NS-NS and R-R plane wave backgrounds,''{\it JHEP} {\bf 0204} (2002) 021, 
hep-th/0202179. 

\bibitem{Dasgupta:1997pu}
K.~Dasgupta and S.~Mukhi,
``BPS nature of 3-string junctions,''
Phys.\ Lett.\ B {\bf 423}, 261 (1998)
[arXiv:hep-th/9711094].

\bibitem{Sen:1997xi}
A.~Sen,
``String network,''
JHEP {\bf 9803}, 005 (1998)
[arXiv:hep-th/9711130].


\bibitem{junc-in-progress}
D. Sadri, ``Giant String Junctions,'' {\it To Appear.}

\end{thebibliography}
\end{document}